\documentclass[conference, a4paper]{IEEEtran}

\usepackage{cite}
\usepackage{amsmath,amssymb,amsfonts}
\usepackage{algorithmic}
\usepackage{graphicx}
\usepackage{textcomp}
\usepackage{xcolor}
\usepackage[caption=false,font=footnotesize]{subfig}
\usepackage{epstopdf}
\def\BibTeX{{\rm B\kern-.05em{\sc i\kern-.025em b}\kern-.08em
    T\kern-.1667em\lower.7ex\hbox{E}\kern-.125emX}}

\begin{document}

\title{Predicting the Mumble of Wireless Channel with Sequence-to-Sequence Models\\
}

\author{\IEEEauthorblockN{Yourui Huangfu, Jian Wang, Rong Li, Chen Xu, Xianbin Wang, Huazi Zhang, Jun Wang}
\textit{Huawei Technologies Co., Ltd.}\\
Hangzhou, China \\
\{huangfuyourui,wangjian23,lirongone.li,xuchen14,wangxianbin1,zhanghuazi,justin.wangjun\}@huawei.com}

\maketitle

\begin{abstract}
Accurate prediction of fading channel in the upcoming transmission frame is essential to realize adaptive transmission for transmitters, and receivers with the ability of channel prediction can also save some computations of channel estimation. However, due to the rapid channel variation and channel estimation error, reliable prediction is hard to realize. In this situation, an appropriate channel model should be selected, which can cover both the statistical model and small scale fading of channel, this reminds us the natural languages, which also have statistical word frequency and specific sentences. Accordingly, in this paper, we take wireless channel model as a language model, and the time-varying channel as talking in this language, while the realistic noisy estimated channel can be compared with mumbling. Furthermore, in order to utilize as much as possible the information a channel coefficient takes, we discard the conventional two features of absolute value and phase, replacing with hundreds of features which will be learned by our channel model, to do this, we use a vocabulary to map a complex channel coefficient into an ID, which is represented by a vector of real numbers. Recurrent neural networks technique is used as its good balance between memorization and generalization, moreover, we creatively introduce sequence-to-sequence (seq2seq) models in time series channel prediction, which can translates past channel into future channel. The results show that realistic channel prediction with superior performance relative to channel estimation is attainable.
\end{abstract}

\begin{IEEEkeywords}
channel prediction, word embedding, prediction diversity
\end{IEEEkeywords}

\section{Introduction}
The fifth generation (5G) wireless communication techniques aim at connecting everything in the physical world as neurons in a neural network. To achieve this goal, 5G systems should be able to interact with environment and ever-increasing numbers of key performance indicators (KPIs) such as user experiences of virtual reality or internet of things should be jointly optimized. When the network optimization problems are becoming more and more complicated, the air interface technology in 5G New Radio (NR) is also facing the same problem with a bunch of new weapons such as massive multiple-input multiple-output (MIMO), non-orthogonal multiple access (NOMA), and Polar/LDPC codes. Fortunately, artificial intelligence (AI) techniques are born for dealing with sophisticated problems and becoming the state of the art in many fields. While many researchers have already used  AI algorithms in air interface blocks, most of air interface technologies are perfectly modeled by information theory, hence their replacements with AI should be carefully checked \cite{you2019ai}. Other than replacement, we believe the AI-enhanced Air Interface (AIeAI) technology will be promising, which combines the benefits of certainty and uncertainty.

In modern radio systems, the most uncertain part of air interface is the wireless channel. The estimated channel quickly become outdated due to rapid channel variation caused by multipath fading, and using the outdated information in transmission will degrade the system performance \cite{duel2007fading, oien2004impact}. Also, precoding of massive MIMO requires excessive channel state information (CSI) feedback, if future CSI is predictable, feedback overhead and pilot resources can be partially saved. However, the wireless channel is a superposition of sinusoids contributed by changing reflectors and scatters,  which is time-varying and intangible, moreover, the channel we can see is estimated channel with estimation error and noises, these problems make the channel prediction hard to realize. Therefore, we need an appropriate channel model, which can not only fit the statistical properties of channel, but also memorize some small scale properties to some extent, as the environment around a base station will not change rapidly, the memorization capability should be helpful. This situation encountered in channel prediction reminds us the similar problems when dealing with languages, in language modeling, a word has a statistical word frequency, meantime, word in different sentences can have different meanings. In this paper, we take wireless channel model as a language model, and the time-varying channel as talking in this language, while the realistic noisy estimated channel can be compared with mumbling. Accordingly, to predict the mumble of wireless channel, is to do channel prediction under a noisy situation with a language modeling way.

To the best of our knowledge, in previous channel prediction works, only one or two features are extracted from a complex channel coefficient, two features are normally the real part and imaginary part or absolute value and phase value, while some researchers focus only on the absolute value \cite{duel2000long} or phase value. Though the power of channel coefficients is the only interest for some applications, it's beneficial to predict complex-valued channel coefficients and obtain power from the squared magnitude of the complex value \cite{ekman2002prediction}. And models used in previous channel prediction methods are various, such as auto-regressive (AR) model \cite{sternad2003channel}, sinusoidal model \cite{chen2006new}, complex-valued neural networks \cite{ding2014fading}, deep neural networks \cite{luo2018channel} and so on. However, before considering the various models, this question should be answered, is it sufficient to extract only two features from a complex channel coefficient?  We believe not. The features should be more and learnable, so that it can capture the small scale properties while the long term properties are kept. In language modeling, word embedding \cite{bengio2003neural} or word vector is the most popular technique, word embedding is a vector of numbers, which can capture context of the word and its relation with other words from a corpus. So can we take a channel coefficient as a word and doing the same embedding process? It is simple, to do this, we can build a vocabulary to map a complex channel coefficient into an ID, which is represented by a vector of real numbers. This innovative method makes the expressive power of our channel model large enough to memorize and fit the realistic channel.

Recurrent neural networks (RNNs) is good at solving sequential problems. For prediction task, while one RNN can only train a sequence in unidirectional way because future elements should not be seen, sequence-to-sequence (seq2seq) models \cite{sutskever2014sequence} can realize bi-directional training by introducing two RNNs for encoder and decoder respectively. Normally, seq2seq models are trained to translate sequence from one domain (e.g., sentences in English) to sequence in another domain (e.g., voices in Mandarin), while in this paper, we creatively exploit seq2seq models in the channel prediction task, so that the encoder side is kept in the past and can be bi-directional trained, and the decoder side is channel in the future. The main contributions of this paper are as follows:

\begin{itemize}
\item The first proposal and implementation of channel prediction algorithms with hundreds of features representing each complex channel coefficient instead of conventional two features.
\item The first demonstration of using seq2seq models and its variants in time series channel prediction. It turns out that the encoder and decoder of seq2seq models with different lengths  can be perfect containers for different time spans of past and future signals.
\item The numerical results with simulation and realistic data indicate the channel prediction model is reliable and robust, and realistic channel prediction with superior performance relative to channel estimation is attainable by using firstly proposed prediction diversity technique.
\end{itemize}

The rest of this paper is organized as follows. The algorithms for channel predictor are presented in Section ~\ref{section:predictor}. The modeling and predicting method is introduced in Section ~\ref{section:modeling}. This is followed by numerical results and discussions in Section ~\ref{section:results}. In Section ~\ref{section:conclusions}, conclusions are given.

\section{CHANNEL PREDICTOR}
\label{section:predictor}
Learning a language model with neural networks is popular for modern natural language processing (NLP) tasks. Normally, a vocabulary should be extracted from the corpus to be learned, which comprises all high-frequency words in this corpus. In a RNN-based language model learning, the conditional probability of each word is computed after all the previous words passing through the RNN cell, where Long Short-Term Memory (LSTM) or Gated Recurrent Unit (GRU) is typically implemented. Corresponding to each word in the vocabulary, a word embedding representing this word is made up of a vector of numbers, which are gradually changing during training until the meaning of this word is encoded in these numbers. Due to the high dimensional space of the vector, a word embedding can carry much more information than a single word.

As we see the noisy estimated channel as mumbling with channel language, a vocabulary of this language should also be extracted from the channel. However, the combinations of amplitude and phase of channel coefficients are infinite, which means the wireless channel only obey a statistical distribution. Fortunately, the Channel Changes (CC) is finite with a practical precision, if we can predict CC from the past, we can predict future channel. Channel coefficients in a past time span can be expressed as $h(t-M:t-1)$, which contains the past $M$ sampled channel, its prediction target is the future $N$ samples, i.e., $\hat{h}(t:t+N-1)$. In this work, CC are calculated in the first place, CC sequence can be expressed as $h'(t-x)=h(t-x)-h(t-x-1)$, where $x=1:M-1$, then $h'$ is used to predict $\hat{h'}$ with future CC. At time $t$, as $\hat{h'}(t) = \hat{h}(t)-h(t-1)$, and $h(t-1)$ is known information at past, $\hat{h}(t)$ can be obtained with $\hat{h}(t) = h(t-1)+\hat{h'}(t)$. Analogously, the $(t+y)^{th}$ sample in the future can be calculated by $\hat{h}(t+y) = h(t-1)+\sum_{k=0}^{y} \hat{h'}(t+k)$ with predicted $\hat{h'}$, where $y=0:N-1$.

As each CC in $h'$ can be seen as a word, we introduce Vocabulary of Channel Changes (VCC) to map $h'$ into IDs. VCC includes top $X$ frequently appearing CC in $h'$ while the rest $L$ CC are out of vocabulary. By considering the numerical precision of $h'$, the size of data, the expressive power of model and GPU memories, an appropriate $X$ should be chosen. Small $X$ makes fitness inaccuracy while large $X$ brings slow convergence speed. Small $L$ introduces interference while large $L$ leads to too many unknown predictions. In this work, $X\approx2000$ and $L\approx500$ are chosen while length of $h'$ is in the order of tens to hundreds of millions. In VCC, an unique ID (usually an integer) is assigned to each CC, the embedding of these integers instead of channel coefficients are inputs of neural networks. Table ~\ref{tab:VCC} shows the top 10 most and least frequently occurring CC and their corresponding IDs in a VCC extracted from a realistic measured channel, where $X$ is the size of this vocabulary. From the top 10 most frequently occurring CC and their frequencies, we can observe pairs of CC, such as $'+0.02-0.02i'$ and $'-0.02+0.02i'$, each pair of CC is symmetrical about the origin or axis, which indicates a specific statistical distribution this wireless channel obeys. In this VCC, CC with occurring frequency higher than 10 are saved, so that the least frequency of CC here is 11. An `unk' token is usually added to vocabulary whose ID is zero for example, so that all occurrences of out-of-vocabulary CC can be replaced with this `unk' token. As mentioned above, every CC has a $e$-length embedding representing it, where $e$ is around 400 in this work, which gives each channel coefficient 400 features. Therefore, number of parameters utilized to represent this channel model is $X \times e \approx 2000 \times 400 = 800k$ without considering the weights in neural networks. It is noteworthy that this work provides a fundamental method to greatly enlarge the expressive power of channel predictor, the model used here is over-parameterized and the ideas such as model pruning can be implemented for practice to speed up the computation. By looking up the VCC, each CC in $h'$ is replaced by its corresponding ID,  then $h'$ is transformed into a new sequence with integers, which can be fed into neural networks for training and predicting. The predicted integers are then transformed back into future $\hat{h'}$ using VCC. With $\hat{h'}$, $\hat{h}$ in future can be obtained.

\begin{table}[!t]
\caption{AN EXAMPLE OF CC WITH TOP10 MOST AND LEAST FREQUENCIES IN VCC}
\begin{center}
\begin{tabular}{|c|c|c|c|c|c|}
\hline
\multicolumn{2}{|c|}{\textbf{Top10 most in VCC}} & & \multicolumn{2}{|c|}{\textbf{Top10 least in VCC}} & \\
\hline
\textbf{ID}&\textbf{CC}&\textbf{Freq.}&\textbf{ID}&\textbf{CC}&\textbf{Freq.} \\
\hline
1&'+0.02-0.02i'&538211&X-9&'+0.2-0.05i'&12 \\
\hline
2&'-0.02+0.02i'&536925&X-8&'-0.04+0.2i'&12 \\
\hline
3&'-0.02-0.02i'&535761&X-7&'-0.03+0.2i'&12 \\
\hline
4&'+0.02+0.02i'&534726&X-6&'-0.1+0.2i'&12 \\
\hline
5&'-0.02+0.01i'&373125&X-5&'-0.06-0.2i'&11 \\
\hline
6&'-0.01+0.02i'&371946&X-4&'-0.05-0.2i'&11 \\
\hline
7&'+0.01+0.02i'&371856&X-3&'+0.01+0.2i'&11 \\
\hline
8&'-0.02-0.01i'&371778&X-2&'+0.2+0.01i'&11 \\
\hline
9&'-0.01-0.02i'&371682&X-1&'+0.05+0.2i'&11 \\
\hline
10&'+0.01-0.02i'&371673&X&'+0.2+0.05i'&11 \\
\hline
\end{tabular}
\label{tab:VCC}
\end{center}
\vspace{-0.8cm}
\end{table}

\section{MODELING AND PREDICTING}
\label{section:modeling}
\begin{figure}[!t]
\setlength{\abovecaptionskip}{0pt}
\setlength{\belowcaptionskip}{0pt}
\centerline{\includegraphics[width=3.5in]{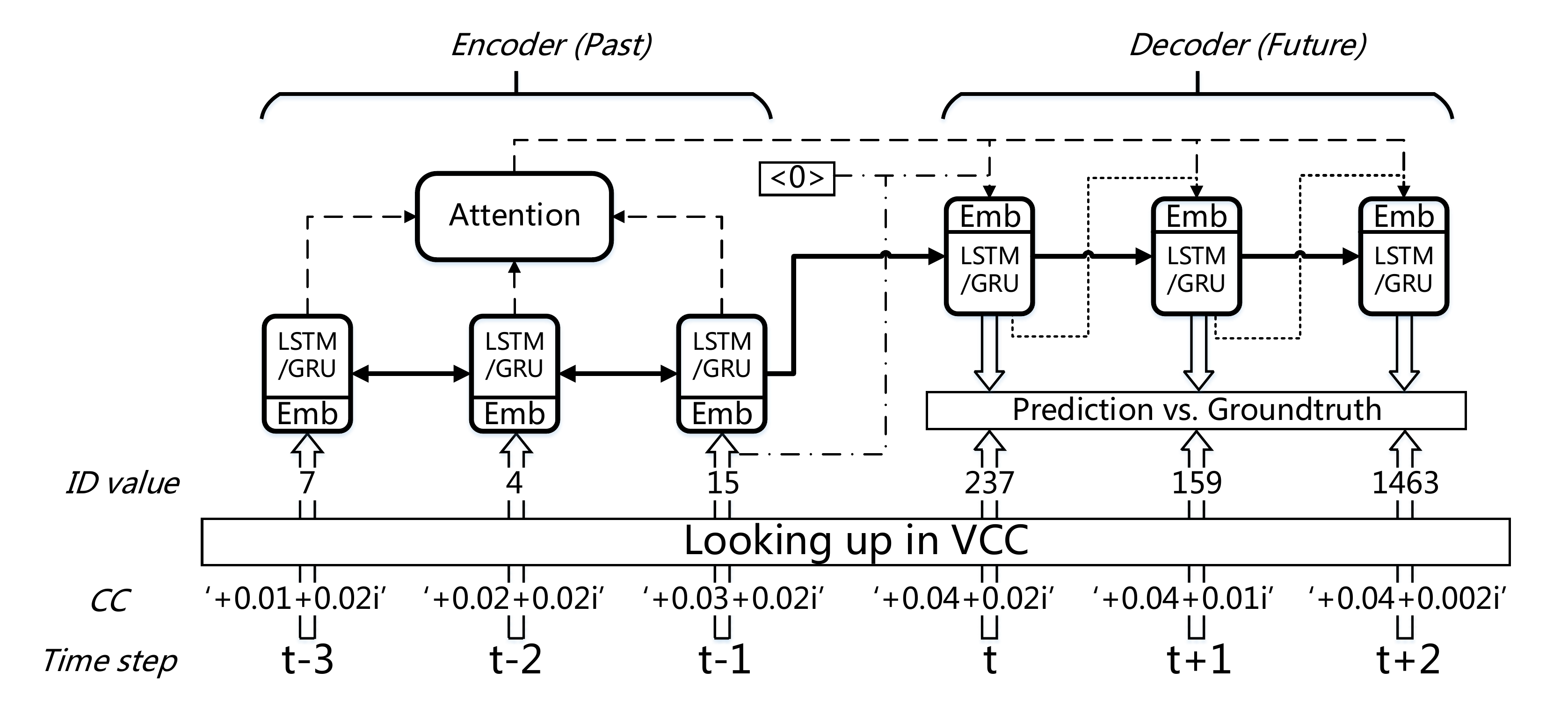}}
\caption{Block diagram of seq2seq model based channel predictor.}
\label{fig:block}
\vspace{-0.1cm}
\end{figure}

To achieve the purpose of predicting the $N$ following CC given the $M$ preceding CC, two solutions are proposed. The natural language generation (NLG) solution uses one RNN combined with backpropagation through time (BPTT) algorithm while the neural machine translation (NMT) solution uses seq2seq models comprising two RNNs, one for encoder and another one for decoder. Normally, these two RNNs belong to different domains, e.g., two kinds of languages, that means they should have different vocabularies. However, in our solution, we use the same vocabulary for both RNNs as they are modeling the same channel model, while encoder is in past and decoder is in future. In figure ~\ref{fig:block}, a block diagram is shown to explain the NMT based channel predictor. At the encoder side, time series with integer IDs $(7, 4, 15)$, representing their CC counterparts in VCC, are fed into the reusable RNN cell, which consists an embedding (EMD) layer and stacked LSTM or GRU networks, the embedding layer contains the embedding vectors for all IDs. The hidden states (solid lines) are transferred through time steps in forward and backward direction, this bidirectional variant can provide better understanding of signals but should only stay at encoder side,  because time order in future cannot be reversed. Normally, only the last hidden state is forwarded to decoder, while in seq2seq models with attention variant, intermediate hidden states with different attention values (weights) are also sent to decoder (dashed lines). Apparently, the attention and bidirectional variants should work together to achieve the largest performance gain. Except the hidden states input, IDs at the last time step of encoder should also be sent to decoder (dash-dotted lines), alternatively, forwarding number zero also works well as the hidden states already possess enough information. The output of decoder is a predicted sequence, by comparing it to ground truth sequence, loss or prediction error can be obtained. It is worth noting that the lengths of input and output sequence are variable and can be different. Furthermore, data amount of train set can be largely increased by sliding time series with a window. For example, the input can be $(4, 15, 237)$ when output is $(159, 1463)$ by sliding the original data with one time step,this is similar to the data augmentation idea in Convolutional Neural Network (CNN) \cite{krizhevsky2012imagenet}.

\begin{figure}[!t]
\setlength{\abovecaptionskip}{0pt}
\setlength{\belowcaptionskip}{0pt}
\centerline{\includegraphics[width=3.5in]{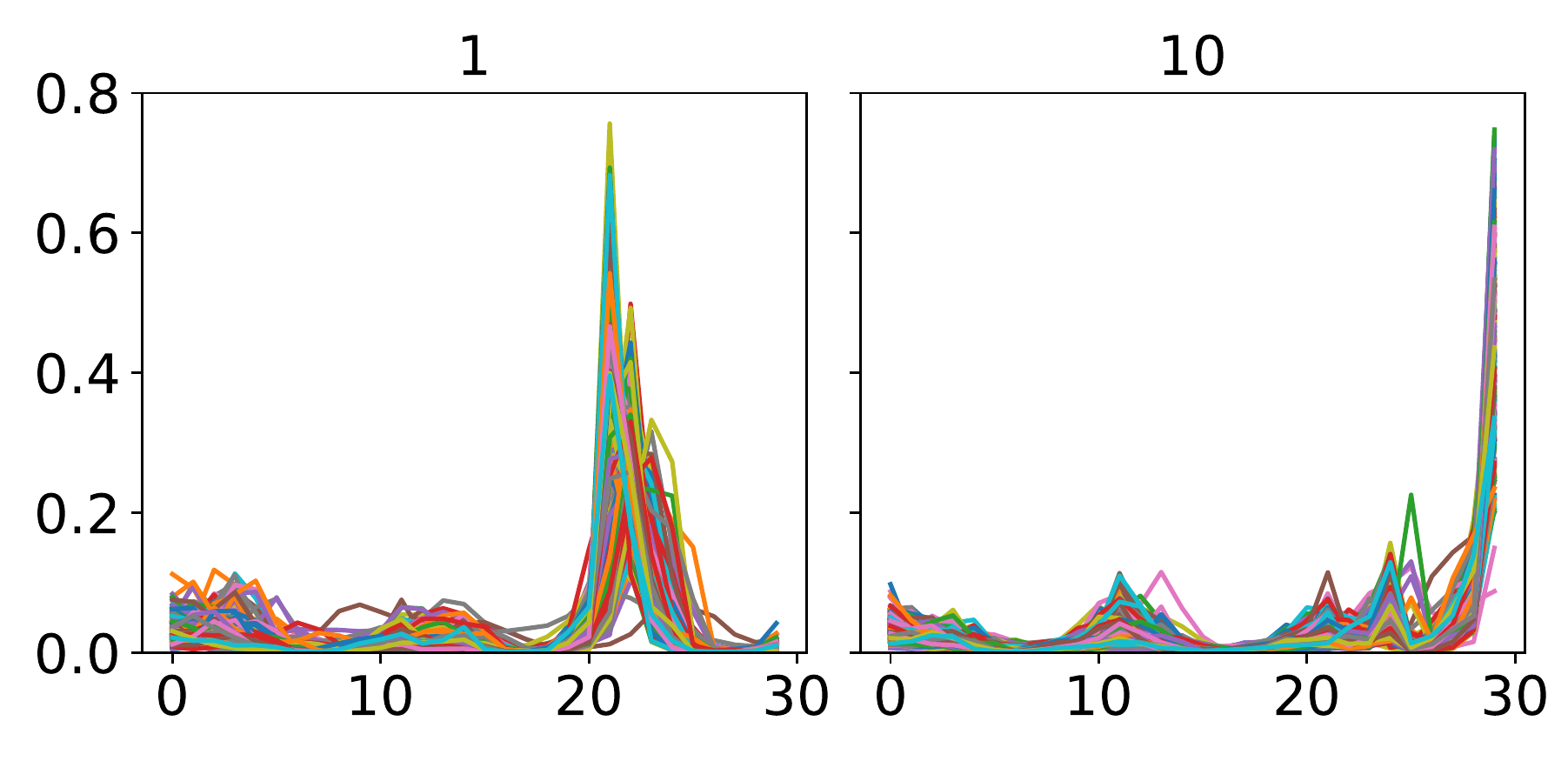}}
\caption{Attentions of 1$^{st}$ and 10$^{th}$ predicted CC on the preceding 30 CC.}
\label{fig:attention}
\vspace{-0.3cm}
\end{figure}

Figure ~\ref{fig:attention} shows the attention values of 1$^{st}$ and 10$^{th}$ predicted CC on the preceding 30 CC for hundreds of NMT based predictions with a well-trained network (route 10 in indoor experiment in Section ~\ref{subsection:measurement}). Apparently, we can tell that the main attention of the first predicted CC focuses on the intermediate hidden state instead of the last or the first hidden state of preceding CC, which means that in bidirectional mode, the hidden state calculated from part of forward direction and other part of backward direction may contribute more valuable information than full forward or full backward. The idea of introducing attention values is similar to the application of autocorrelation values in linear prediction method based on the AR modeling, only this attention mechanism is more powerful because the attention target is a hidden state of neural network while the autocorrelation target is just a channel coefficient. Observed from this figure, attention values through different tests are similar, that is essential for a robust channel prediction model. For the 10$^{th}$ predicted CC, the main attention is on the last hidden state, which means the latest signal is more and more important when predicting samples further ahead.

To simplify, $M:N$ is used to represent the operation of utilizing $M$ preceding CC to predict $N$ following CC, in practice, this $M:N$ prediction operation should be repeated to ensure the prediction accuracy as the prediction error will always accumulate in time. Though predicting channels infinitely into the future is impossible, a better prediction algorithm can extend $N$ under a limited prediction error. In channel prediction task, regarding the time interval between sampled channel or Doppler shift, the unit of $M$ and $N$ can also be time span or wavelength. For example, task in figure ~\ref{fig:attention} can be expressed as a 30:10 samples channel prediction task, as the time interval between sampled channel is 1 millisecond (ms), it's also a 30:10 ms channel prediction task. As the user speed in this case is 3km/h, the highest Doppler frequency is about 10Hz at center frequency 3.45GHz, the distance traveled during the prediction can be measured in wavelengths $W$,

\begin{equation}
W = Tf_d=\frac{Tv}{\lambda}=\frac{Tvf_c}{c}
\label{eq:wavelength}
\end{equation}

where $T$ is prediction time span, $f_d$ is doppler frequency, $v$ is user velocity in $m/s$, $\lambda$ is carrier wavelength, $f_c$ is the carrier frequency, and $c$ is the speed of light. By calculating equation ~\ref{eq:wavelength}, the previous task can be expressed as a 0.3:0.1 wavelengths in space.

In this paper, each RNN cell has a two-layered LSTM/GRU with hidden layers of size 1000. For fair comparison, we train 2 epochs for every model with Adam as optimizer and same annealing recipe of learning rate. After training, the predictor is tested with an M:N predicting operation, then every $N$-length segmented data in testing channel sequence $h_{t}$   is replaced with $N$ predicted samples, so that a new predicted channel sequence $\hat{h}_{t}$ is obtained. The normalized mean square error (NMSE) of the prediction can be expressed as equation ~\ref{eq:NMSE},

\begin{equation}
NMSE = \frac{E\{\|h_{t}-\hat{h}_{t}\|^{2}_{F}\}}{E\{\|\hat{h}_{t}\|^{2}_{F}\}}
\label{eq:NMSE}
\end{equation}

where $\|\cdot\|_{F}$ is Frobenius norm. For radio system, what really matters is the final throughput. Therefore, in some of the following experiments, block error rate (BLER) performance is also calculated with the predicted channel coefficients.

\section{NUMERICAL RESULTS AND DISCUSSIONS}
\label{section:results}
Performance evaluation of this channel predictor is carried out on two scenes: link level simulation (LLS) in section ~\ref{subsection:simulation} and realistic measured scene in section ~\ref{subsection:measurement}. Orthogonal Frequency Division Multiplexing (OFDM) system with 20MHz bandwidth and QPSK modulation is used for both scenes. All the data and results of LLS presented here are based on the tapped-delay line (TDL) channel model \cite{meredith2016study}, which takes care of mmWave propagation ranging from 6 to 100GHz. Specifically, TDL-C is designed for non-line-of-sight (NLOS) propagation, in this paper a long delay spread 300ns is set in the TDL-C model and noise-free channel coefficients on 31 propagation paths are introduced for training. For realistic measured scene, indoor and outdoor measurements are carried out at 3.45GHz carrier frequency.

\subsection{Simulation}
\label{subsection:simulation}
\begin{figure}[!t]
\setlength{\abovecaptionskip}{0pt}
\setlength{\belowcaptionskip}{0pt}
\centering
\subfloat[]{\includegraphics[width=3.5in]{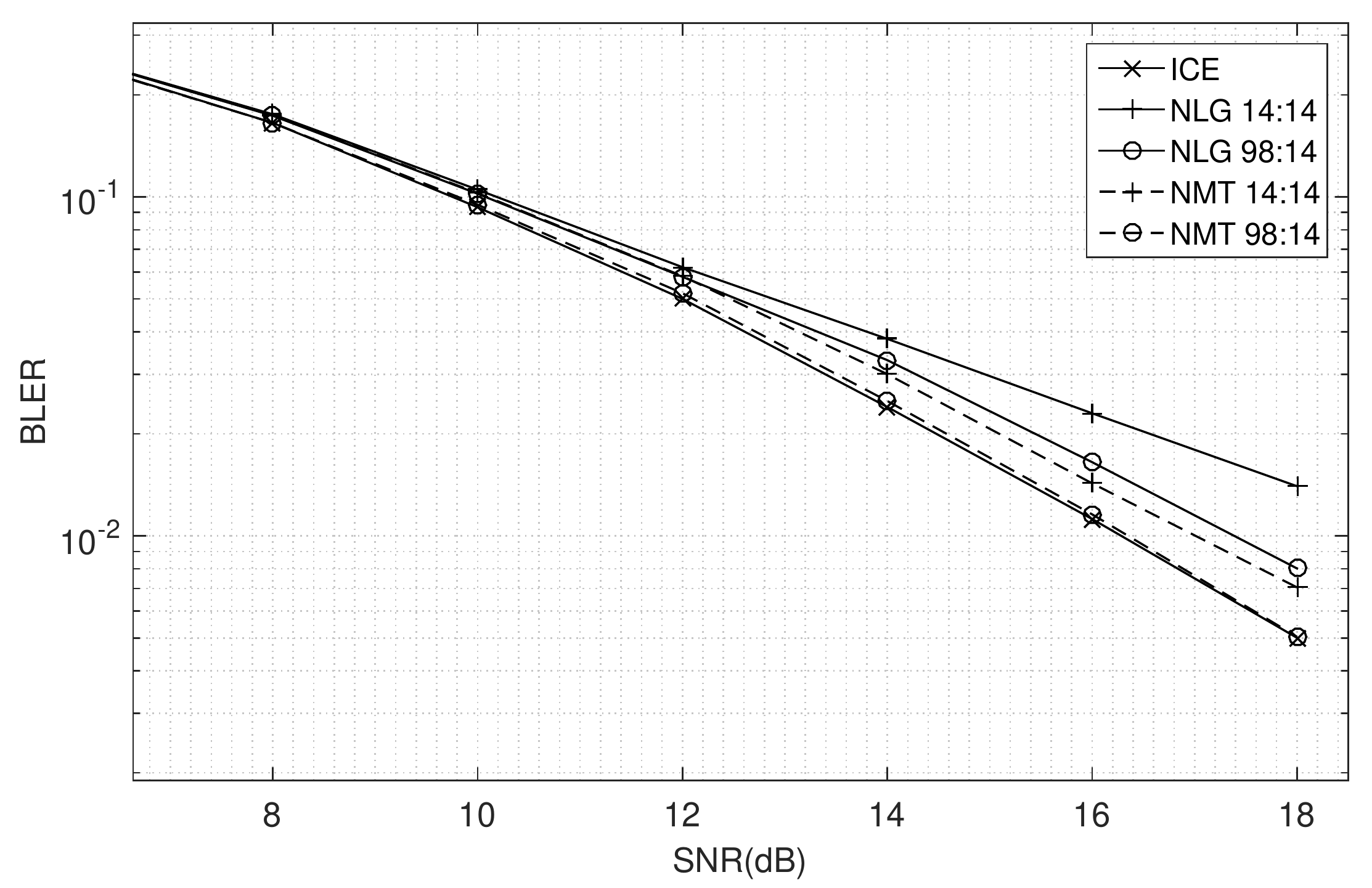}
\label{fig:NLGa}}\\
\subfloat[]{\includegraphics[width=3.5in]{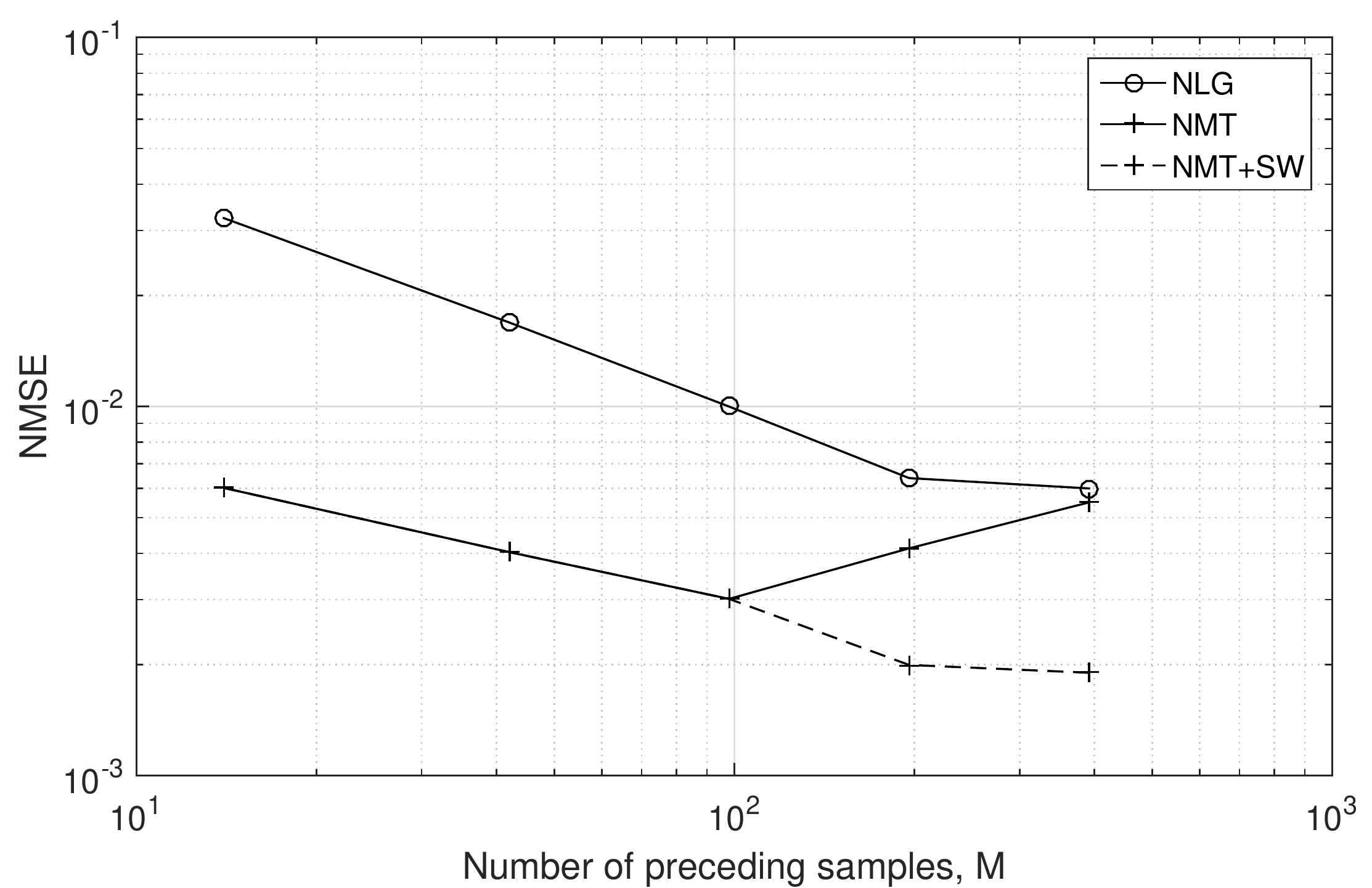}
\label{fig:NLGb}}
\caption{(a) BLER performance of decoding under NLG and NMT solutions with varying $M$ at $N=14$ symbols (b) NMSE vs. $M$ at $N=14$ symbols for NLG, NMT and NMT combined with sliding window (SW) technique.}
\label{fig2}
\vspace{-0.4cm}
\end{figure}

Channel predictor is investigated for LLS to avoid the impact of noise, interference and hardware imperfections. The NLG solution and NMT solution are compared at user speed 100km/h, the unit of $M:N$ is symbol, and 14 symbols equal to one millisecond. In this case, train set is channel impulse response (CIR) from 31 paths in time domain for 100 seconds, while test set is obtained from another 10 seconds with a different random seed. In figure ~\ref{fig:NLGa}, performance of decoding with ideal channel coefficients, i.e., ideal channel estimation (ICE), is shown as a baseline, NLG and NMT solution with varying $M$ are investigated. When $N$ is fixed to 14 symbols, larger $M$ introduces more preceding symbols for learning and predicting. From figure ~\ref{fig:NLGa}, we can tell that by increasing $M$, better performance can be achieved, from figure ~\ref{fig:NLGb}, it is shown that the performance gains result from more accurate prediction according to NMSE value. However, reducing prediction error by increasing $M$ will finally saturate at $M=Ms$, which is around 30 samples shown in previous work \cite{duel2007fading}. Based on the neural-network algorithm we used, the memory span of this channel predictor is six times larger with 196 samples derived from NLG solution.

Normally, the saturate point shows fitting and memorizing capability of the model together with its hyperparameters, while for data-driven models, the size of train set may also limit this capability, to break the bottleneck, a larger dataset should be given. In figure ~\ref{fig:NLGb}, NMT solution trained with the same data set as NLG solution is shown as solid line with plus-sign markers, we can see that when $M$ is larger than 98, prediction is getting worse. It means the same dataset is not enough to define this NMT model with large $M$. Fortunately, the time series sequence is continuous, if we slide the original data set with a time offset, i.e., sliding window technique, an augmented data set is obtained. Though this new data set is all from the old sequence, it's beneficial for NMT based channel predictor to understand that $M$ and $N$ samples are all from a continuous long sequence. In this experiment, a sliding window is chosen to make the date set five times larger, and the prediction result is shown as dashed line, we can see that prediction error is going down at 196 samples and nearly saturate under this amount of data set. In the comparison of NMT and NLG solutions, NMT solution with 14:14 symbols outperforms all NLG solutions on BLER performance and NMSE value, even more, NMT solution with 98:14 symbols can reach the ICE performance bound. It means the prediction error of 98:14 symbols NMT solution is below the critical value for perfectly decoding with imperfect channel coefficients \cite{zhou2004accurate}. Compared to NLG channel prediction, the fitting ability of NMT solution is much stronger.

\begin{figure}[!t]
\setlength{\abovecaptionskip}{0pt}
\setlength{\belowcaptionskip}{0pt}
\centerline{\includegraphics[width=3.5in]{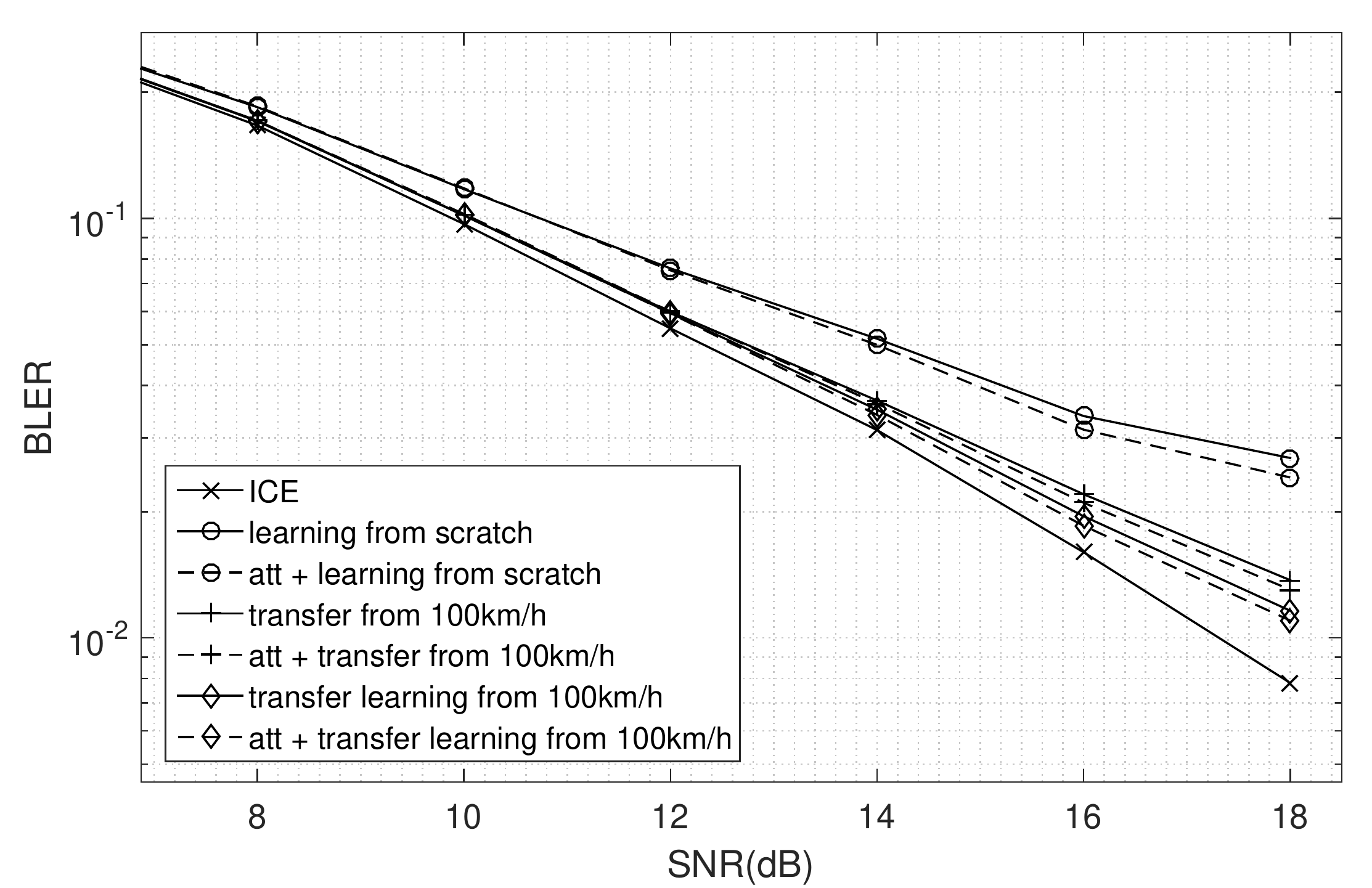}}
\caption{BLER performance of decoding with 30:30 ms NMT predicted channel at 3km/h user speed, predicted channel is obtained with learning from scratch, directly transfer, transfer learning and their attention variants.}
\label{fig:transfer}
\vspace{-0.5cm}
\end{figure}

Transfer learning is efficient when solving a problem while the knowledge of a related problem has already been learned. As we have already learned channel model at 100km/h user speed, a slower or faster time-variant channel should be efficiently predicted or learned by transferring knowledge. In this experiment, we work on a channel with 3km/h user speed, as the channel changes slower, we can predict a longer time span with $M \times S : N \times S$ where S is the sampling rate and the channel coefficients sequence is sampled followed by calculation of $h'$. It is worth noting that when using transfer learning, VCC used to transform $h'$ should be the same as VCC used by the model being transferred. In this case, $S=30$ and $M=N=14$ symbols are applied, that is, a 30:30 ms NMT solution for channel prediction. The predicted $\hat{h'}$ can be recovered to original time interval by interpolating operation. As a result of the sampling operation, train set is 30 times smaller than original data set. For data-driven models, the size of data set is essential, which we have already proved in the previous experiment. If we learn from scratch with this small train set, even after 20 epochs training, around 1dB performance loss at $BLER=0.1$ is shown in figure ~\ref{fig:transfer}. However, if the well-trained 100km/h channel model is directly used to predict sampled 3km/h channel, 0.8dB performance loss can be reduced. From these results, we can see that even without learning, predicting slowly-varying channel with rapidly-varying channel model is feasible. With transfer learning, the performance is further improved, where the 100km/h channel model is being modified to fit the sampled 3km/h channel according to the input data. Moreover, attention variant of seq2seq models can be utilized and further performance gains are achievable as shown in figure ~\ref{fig:transfer}. It is worth noting that for higher SNR, the performance gap between prediction and estimation is larger, which is reasonable because estimation error become smaller and prediction error dominates the performance. However, for cases such as adaptive transmission, estimated channel is outdated, performance difference should be compared between predicted channel and outdated channel, where predicted one should be much better. In this paper, BLER performance is compared only for more directly understanding.

\subsection{Measurement}
\label{subsection:measurement}
\begin{figure}[!t]
\setlength{\abovecaptionskip}{0pt}
\setlength{\belowcaptionskip}{0pt}
\centerline{\includegraphics[width=3.5in]{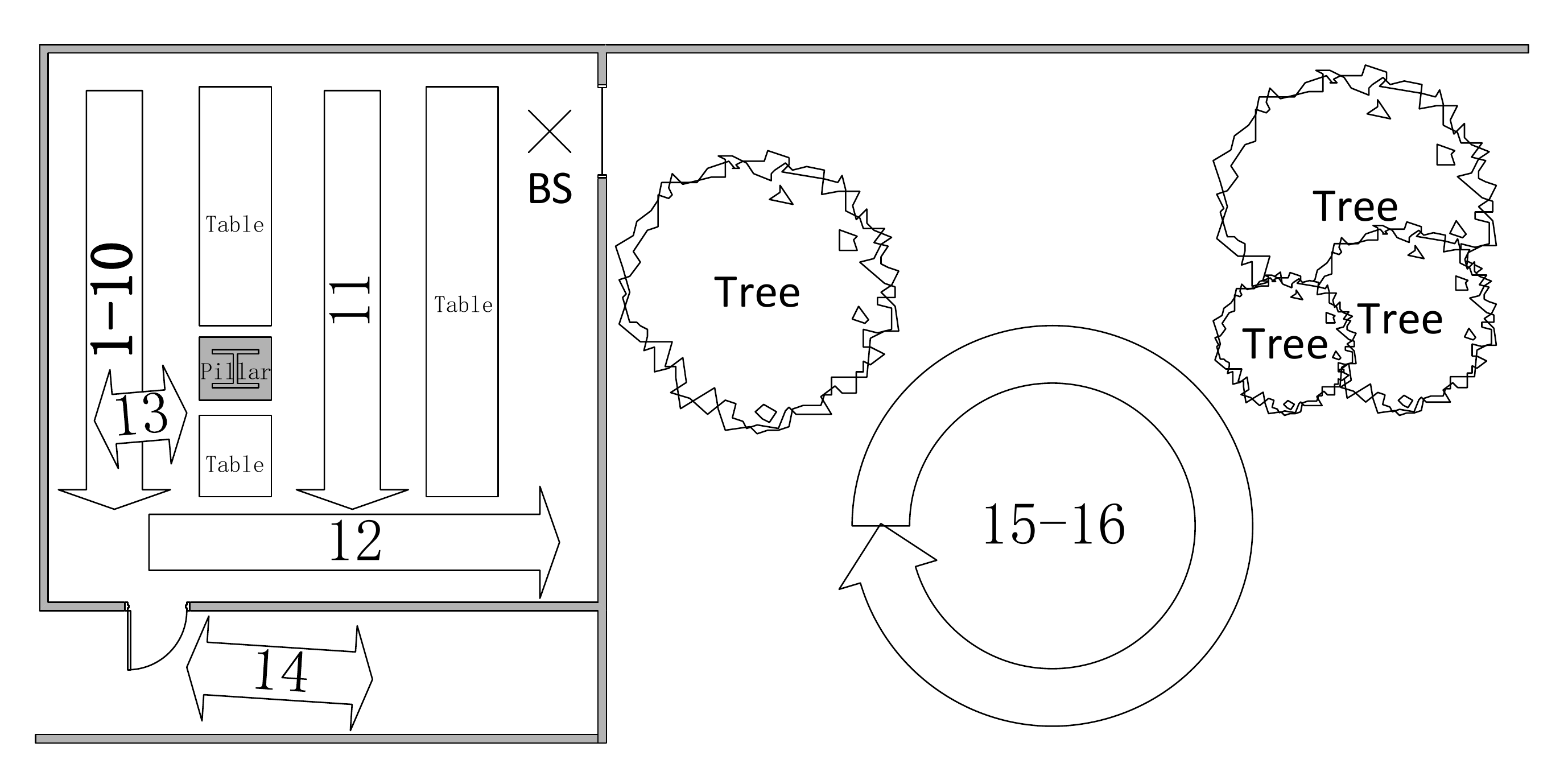}}
\caption{Schematic diagram of environment for indoor and outdoor channel measurement.}
\label{fig:schematic}
\vspace{-0.1cm}
\end{figure}

In realistic scene, measured channel coefficients are forwarding to channel predictor. The indoor and outdoor channel measurements are conducted in an environment shown as a schematic diagram in figure ~\ref{fig:schematic}, which includes a 20m $\times$ 20m room at 4th floor for indoor measurement, a 50m $\times$ 50m garden with surrounding buildings for outdoor measurement, and lines with arrowhead indicating routes and directions of moving user equipment (UE). In the room, floor to ceiling height is 3m and the heights of base station (BS) and UE are 2m and 1m. UE speed for routes 1 to 12 is 3km/h while route 13 is moving around behind pillar, and route 14 is moving around in the corridor. For outdoor scene, BS is pointed to the garden through a window 15 meters above the ground and routes 15 to 16 is moving in circle with 3km/h. UE uses 1 antenna while BS uses 1 antenna for outdoor, 2 antennas for indoor. Different from simulation scene, channel estimation and prediction are in the frequency domain as the accurate sinusoids of propagation channel are difficult to obtain due to hardware limitation and estimation error, therefore, channel frequency response (CFR) is estimated from pilot symbols. The least-square estimated channel is truncated at time domain while significant taps are kept. The time-varying channel from different subcarriers can also be aligned one by one to form one long sequence, which can largely increase the train set.

For indoor scene, channel predictor with attention based 30:10 ms NMT solution is used and the generalization capability of this channel predictor is being investigated. Not only the channel from different subcarriers are trained together, channel from different ports corresponding to 2 transmitting antennas are also trained together, the goal is to generalize the statistical channel model from training on limited route and transferring the knowledge on predicting other routes. Imagine such a scenario, where the AI-enhanced indoor distributed antenna system (IDAS) was trained with channel of an UE in a room and is capable to predict channel of other UEs in other rooms with similar layout. In this experiment, training is carried out with train set composed of data from routes 1 to 9, which are following the same route and are just collected at different times. Data from routes 10 to 14 are directly predicted using the well-trained model, i.e., transferring knowledge directly without transfer learning (see section ~\ref{subsection:simulation}). The prediction error is shown in table ~\ref{tab:indoor}, for comparison, routes 8 and 9 are also predicted with this model they have contributed. We can tell from the table that prediction error of routes 10 to 13 are close to trained routes 8 and 9, moreover, though route 10 is following the same route with train set, route 12 and NLOS route 13 have better prediction accuracy than it, which shows the generalization capability of this model is good. These results also indicate that transferring channel model among different locations, ports and times is feasible. However, route 14 in the corridor seems following different properties of propagation as the NMSE is much larger than the routes inside the room. Fortunately, transfer learning can be easily done on partial data of route 14, it can be seen that the prediction error of the rest of data is reduced. However, to further reduce the prediction error, more data in the corridor should be collected and trained. For a building with AI-enhanced IDAS, it would be easy to collect CSI in all the corridors having antennas in a very short time and generate a large enough data set for training a generalized corridor channel model.

\begin{table}[!t]
\caption{NMSE OF PREDICTION ON ROUTES 8 TO 14 WITH TRANSFERRED KNOWLEDGE FROM ROUTES 1 TO 9 MODEL, TRANSFER LEARNING (TL) IS INVESTIGATED FOR ROUTE 14}
\begin{center}
\begin{tabular}{|c|c|c|c|c|}
\hline
\textbf{Route}&\textbf{8}&\textbf{9}&\textbf{10}&\textbf{11} \\
\hline
\textbf{NMSE}&0.0061&0.0057&0.0062&0.0064 \\
\hline
\textbf{Route}&\textbf{12}&\textbf{13}&\textbf{14}&\textbf{14+TL} \\
\hline
\textbf{NMSE}&0.0058&0.0054&0.0108&0.0075 \\
\hline
\end{tabular}
\label{tab:indoor}
\end{center}
\vspace{-0.1cm}
\end{table}

Compared to indoor scene, BLER performance of the channel predictor for outdoor scene is studied at various signal-to-noise ratio (SNR) values by using an adjustable attenuator. In these experiments, channel predictor with attention based 20:20 ms NMT is used. For comparison, channel sampling rate (CSR) of estimator, i.e., frequency of channel estimation, is designed to be 400Hz or 2kHz. Channel coefficients on 1320 subcarriers estimated with 400Hz CSR are used for training of channel predictor, and the total length of train set is around 0.2 billion, which is from around one minute CSI collection. Routes 15 to 16 are following the same circling route, data from route 15 is used for training and route 16 for testing. It is worth noting that though channel predictor is trained with $N=20$ ms, we still can use this predictor to inference on different time spans, e.g., 10ms or 30ms, only the predictor's understanding of small scale properties will be slightly different.

\begin{figure}[!t]
\setlength{\abovecaptionskip}{0pt}
\setlength{\belowcaptionskip}{0pt}
\centerline{\includegraphics[width=3.5in]{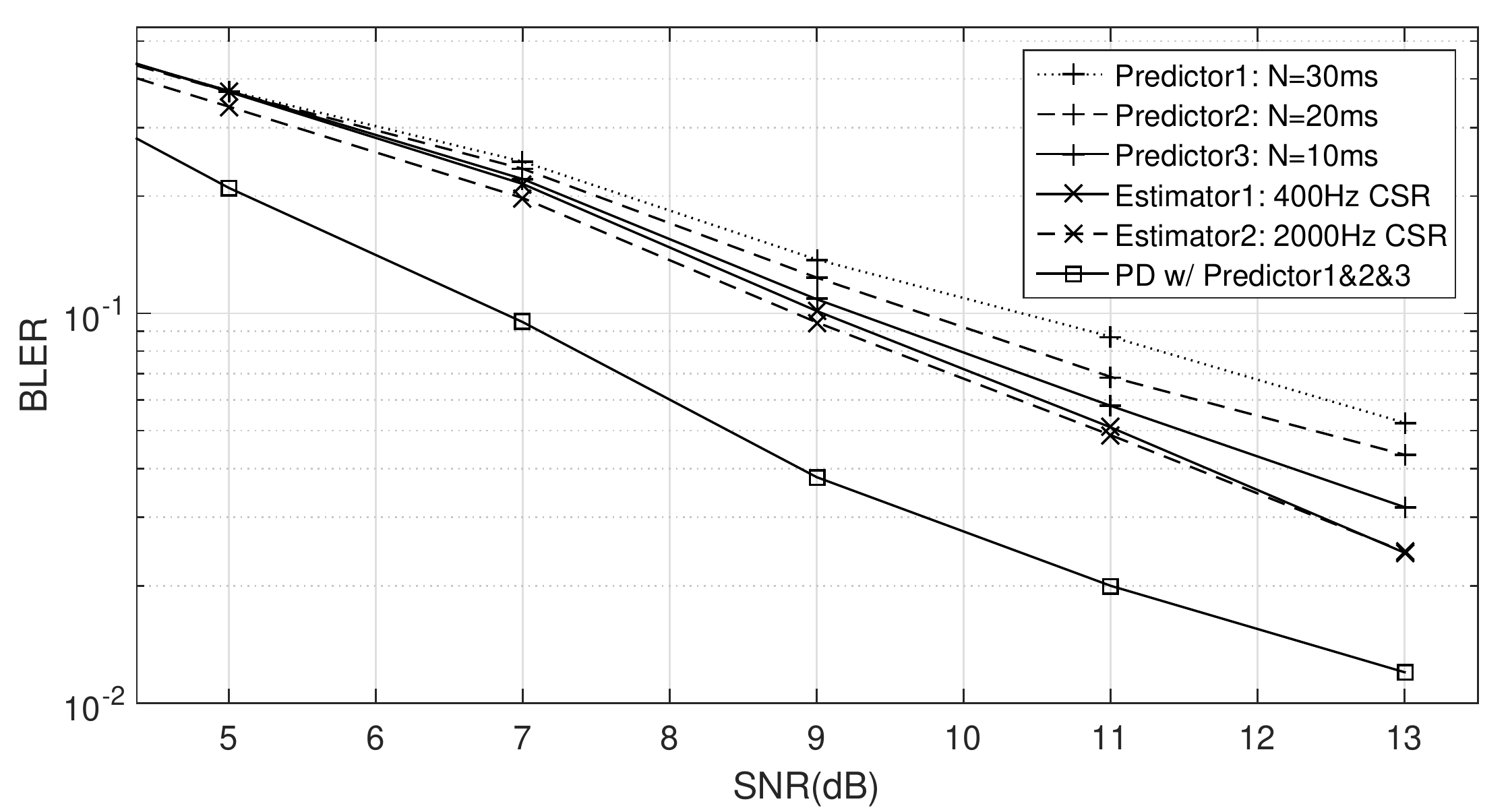}}
\caption{BLER performance of decoding with two estimators and three predictors. Estimator 1 and 2 use 400Hz and 2000Hz CSR respectively, and predictors with varying $N$ are trained on 400Hz estimated channel coefficients when $M$ is 20 ms. Results of prediction diversity (PD) by calculating maximum CFR of all three predictors are shown with around 2dB gain.}
\label{fig:diversity}
\vspace{-0.0cm}
\end{figure}

The decoding performance with predicted and estimated channel coefficients for route 16 is shown in figure ~\ref{fig:diversity}. We can tell that longer prediction time span introduces larger performance loss, and at $BLER=0.1$ the decoding performance of $N=10$ ms predictor is around 0.1dB inferior to decoding performance of estimator with 400Hz CSR. When $M=20$ ms and $N=10$ ms, at least one third pilot or CSI feedback resources can be saved, here is a trade-off between improving BLER performance and saving time-frequency resources, both of them can finally increase the system throughput.

\begin{figure}[!t]
\setlength{\abovecaptionskip}{0pt}
\setlength{\belowcaptionskip}{0pt}
\centerline{\includegraphics[width=3.5in]{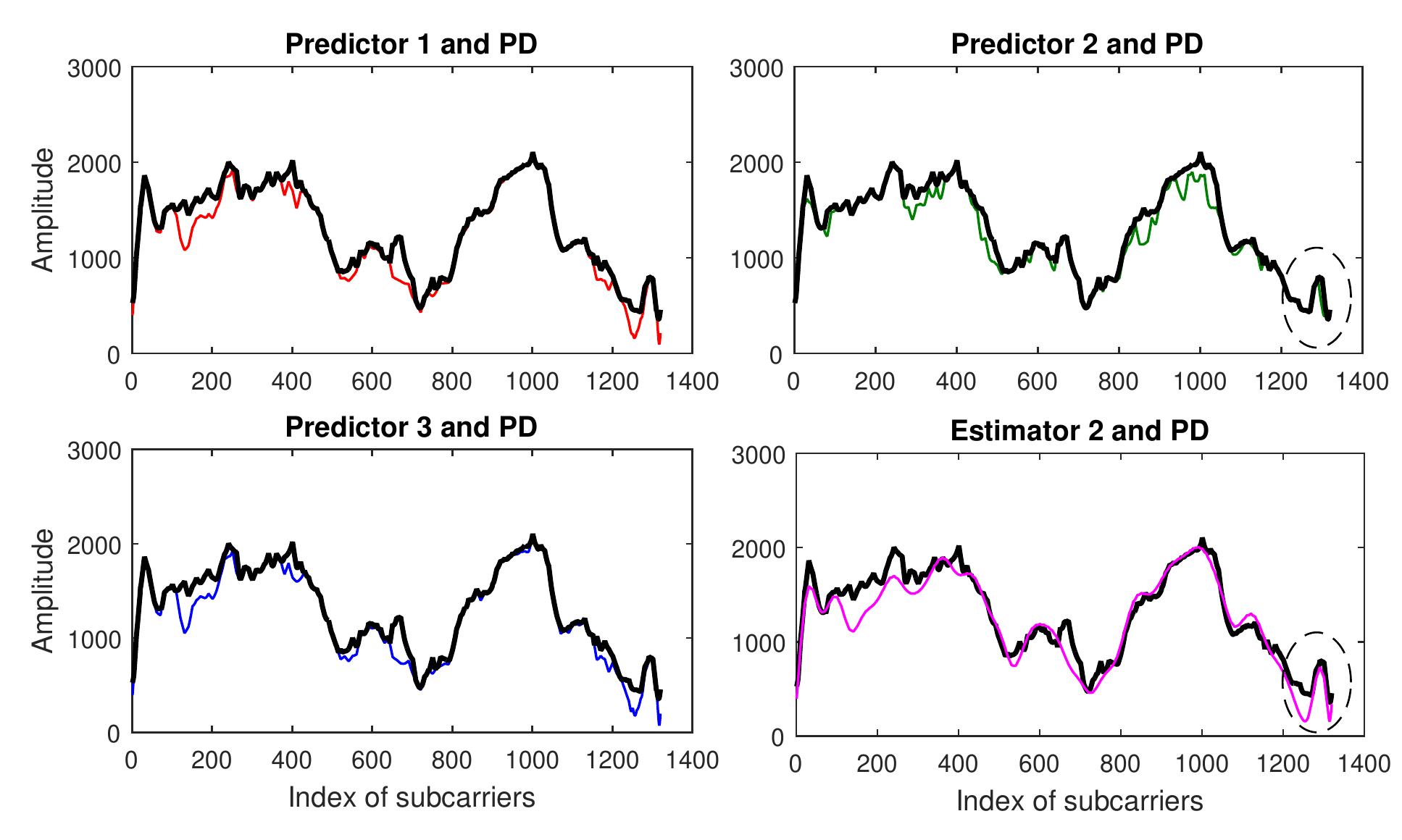}}
\caption{Amplitudes of channel from predictors $1\&2\&3$ and estimator 2 are compared to PD (thicker line) in a random frame, where only channel of predictor 2 and PD can decode it correctly. Deep fades are surrounded by dashed circles.}
\label{fig:amplitude}
\vspace{-0.1cm}
\end{figure}

The second trade-off is between improving BLER performance and saving computing resources. As shown in figure ~\ref{fig:diversity}, improving BLER performance by increasing CSR is not easy due to the inherent estimation error (estimator 2). Fortunately, predictor can outperform estimator with the cost of some computing resources. For example, we can pick the maximum CFR from three predictors for each subcarrier and combine them into a new prediction, as this operation is similar to antenna diversity, we name this new prediction as prediction diversity (PD). In figure ~\ref{fig:amplitude}, channel amplitudes versus subcarriers in a random frame are compared for three predictors, estimator 2 and PD. In this frame, only channel of predictor 2 can decode this frame correctly while other predictors or estimators will fail, though channel predicted by predictor 2 is the most different from channel estimated by estimator 2, the randomness of prediction can provide more possibilities to overcome estimation error. Especially for deep fades surrounded by the dashed circles, where estimation error is largest and the phase features of CFR are seriously influenced by noise, however, predictors with different settings may or may not fall into the deep fades, and when some predictor does not fall into the deep fades, more reasonable phase features can be kept, even it's not the real channel that data go through, it's much better than using a wrong estimated results. By using PD for decoding at each frame, around 2dB performance gain can be achieved as shown in figure ~\ref{fig:diversity}. When improving BLER performance is critical and computing resources are sufficient, predicting more times and combine them with PD technique can be utilized for larger gains.

In this experiment, $M$ is 20 ms, which can be increased for better prediction. However, the third trade-off here is between decreasing prediction error and decreasing block error. The training target is to decrease prediction error while the radio system is willing to decrease the block error. If the channel predictor is accurate enough, performance of this predictor will be nearly the same with estimator, however, its ability of compensating estimation error is also weakened, which means the PD methods will be useless. Therefore, for a specific system, a specific strategy should be designed for channel predictor.

\section{Conclusions}
\label{section:conclusions}
In this paper, we proposed a new channel prediction algorithms with hundreds of learnable features representing each complex channel coefficient instead of conventional two features. The additional features are essential for a channel model to balance fitting the statistical model and memorizing small scale properties. To implement this algorithm, we tried recurrent neural networks as a start, furthermore, seq2seq models and its variants with better performance are proved for time series channel prediction task. It turns out that the encoder and decoder of seq2seq models with different lengths but the same Vocabulary of Channel Changes can be perfect containers for different time spans of past and future signals. The numerical results with simulation and realistic data indicate the channel prediction model is reliable and robust, and realistic channel prediction with superior performance relative to channel estimation is attainable by using firstly proposed prediction diversity technique. These results show us a promising future for AI-enhanced channel prediction.

In future works, the influence of memorizing capability of neural network based channel predictor on small scale fading should be further investigated. More antenna and user equipments are needed. Prediction diversity technique should be verified on various cases. Deal with the difficulties of deployment of this algorithm in a real system.

\bibliographystyle{IEEEtran}
\bibliography{CP1}

\begin{thebibliography}{10}
\providecommand{\url}[1]{#1}
\csname url@samestyle\endcsname
\providecommand{\newblock}{\relax}
\providecommand{\bibinfo}[2]{#2}
\providecommand{\BIBentrySTDinterwordspacing}{\spaceskip=0pt\relax}
\providecommand{\BIBentryALTinterwordstretchfactor}{4}
\providecommand{\BIBentryALTinterwordspacing}{\spaceskip=\fontdimen2\font plus
\BIBentryALTinterwordstretchfactor\fontdimen3\font minus
  \fontdimen4\font\relax}
\providecommand{\BIBforeignlanguage}[2]{{%
\expandafter\ifx\csname l@#1\endcsname\relax
\typeout{** WARNING: IEEEtran.bst: No hyphenation pattern has been}%
\typeout{** loaded for the language `#1'. Using the pattern for}%
\typeout{** the default language instead.}%
\else
\language=\csname l@#1\endcsname
\fi
#2}}
\providecommand{\BIBdecl}{\relax}
\BIBdecl

\bibitem{you2019ai}
X.~You, C.~Zhang, X.~Tan, S.~Jin, and H.~Wu, ``Ai for 5g: research directions
  and paradigms,'' \emph{Science China Information Sciences}, vol.~62, no.~2,
  p. 21301, 2019.

\bibitem{duel2007fading}
A.~Duel-Hallen, ``Fading channel prediction for mobile radio adaptive
  transmission systems,'' \emph{Proceedings of the IEEE}, vol.~95, no.~12, pp.
  2299--2313, 2007.

\bibitem{oien2004impact}
G.~Oien, H.~Holm, and K.~J. Hole, ``Impact of channel prediction on adaptive
  coded modulation performance in rayleigh fading,'' \emph{IEEE Transactions on
  Vehicular Technology}, vol.~53, no.~3, pp. 758--769, 2004.

\bibitem{duel2000long}
A.~Duel-Hallen, S.~Hu, and H.~Hallen, ``Long-range prediction of fading
  signals,'' \emph{IEEE Signal processing magazine}, vol.~17, no.~3, pp.
  62--75, 2000.

\bibitem{ekman2002prediction}
T.~Ekman, ``Prediction of mobile radio channels: modeling and design,'' Ph.D.
  dissertation, Institutionen f{\"o}r materialvetenskap, 2002.

\bibitem{sternad2003channel}
M.~Sternad and D.~Aronsson, ``Channel estimation and prediction for adaptive
  ofdm downlinks,'' in \emph{IEEE Vehicular Technology Conference},
  vol.~2.\hskip 1em plus 0.5em minus 0.4em\relax IEEE; 1999, 2003, pp.
  1283--1287.

\bibitem{chen2006new}
M.~Chen, T.~Ekman, and M.~Viberg, ``New approaches for channel prediction based
  on sinusoidal modeling,'' \emph{EURASIP Journal on Advances in Signal
  Processing}, vol. 2007, no.~1, p. 049393, 2006.

\bibitem{ding2014fading}
T.~Ding and A.~Hirose, ``Fading channel prediction based on combination of
  complex-valued neural networks and chirp z-transform,'' \emph{IEEE
  Transactions on Neural Networks and Learning Systems}, vol.~25, no.~9, pp.
  1686--1695, 2014.

\bibitem{luo2018channel}
C.~Luo, J.~Ji, Q.~Wang, X.~Chen, and P.~Li, ``Channel state information
  prediction for 5g wireless communications: A deep learning approach,''
  \emph{IEEE Transactions on Network Science and Engineering}, 2018.

\bibitem{bengio2003neural}
Y.~Bengio, R.~Ducharme, P.~Vincent, and C.~Jauvin, ``A neural probabilistic
  language model,'' \emph{Journal of machine learning research}, vol.~3, no.
  Feb, pp. 1137--1155, 2003.

\bibitem{sutskever2014sequence}
I.~Sutskever, O.~Vinyals, and Q.~V. Le, ``Sequence to sequence learning with
  neural networks,'' in \emph{Advances in neural information processing
  systems}, 2014, pp. 3104--3112.

\bibitem{krizhevsky2012imagenet}
A.~Krizhevsky, I.~Sutskever, and G.~E. Hinton, ``Imagenet classification with
  deep convolutional neural networks,'' in \emph{Advances in neural information
  processing systems}, 2012, pp. 1097--1105.

\bibitem{meredith2016study}
J.~Meredith, ``Study on channel model for frequency spectrum above 6 ghz,''
  3GPP TR 38.900, Jun, Tech. Rep., 2016.

\bibitem{zhou2004accurate}
S.~Zhou and G.~B. Giannakis, ``How accurate channel prediction needs to be for
  transmit-beamforming with adaptive modulation over rayleigh mimo channels?''
  \emph{IEEE Transactions on Wireless Communications}, vol.~3, no.~4, pp.
  1285--1294, 2004.

\end{thebibliography}
\end{document}